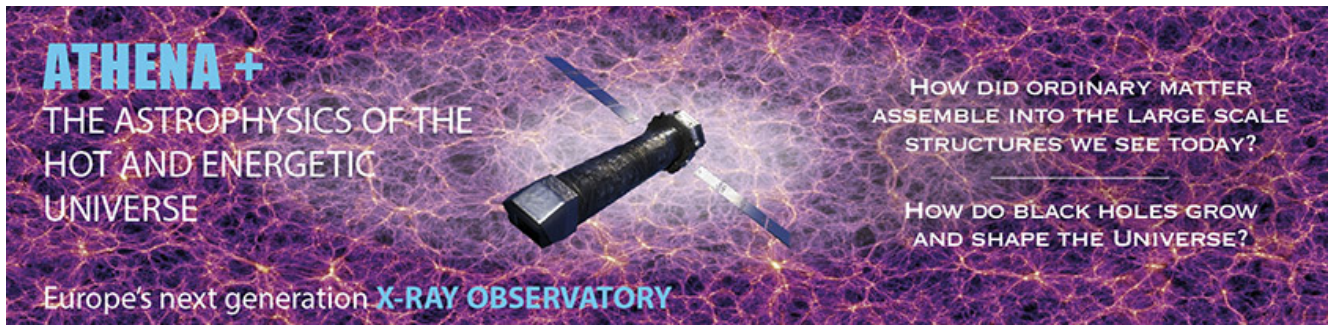

# The Hot and Energetic Universe

An *Athena+* supporting paper

# The astrophysics of supernova remnants and the interstellar medium

## Authors and contributors

**A. Decourchelle, E. Costantini,** C. Badenes, J. Ballet, A. Bamba, F. Bocchino, J. Kaastra, D. Kosenko, R. Lallement, J. Lee, M. Lemoine-Goumard, M. Miceli, F. Paerels, R. Petre, C. Pinto, P. Plucinsky, M. Renaud, M. Sasaki, R. Smith, V. Tatischeff, A. Tiengo, L. Valencic, J. Vink, D. Wang, J. Wilms.

# 1. EXECUTIVE SUMMARY

The study of both supernova remnants and the interstellar medium is essential for understanding the final stages of stellar evolution and their feedback on the evolution of galaxies through injection of energy and heavy elements. These studies are also crucial for understanding the physics of supernovae and their cosmological implication, as well as the origin of galactic cosmic rays.

Supernovae play a major role as the dominant sources of heavy elements in the Universe driving the chemical evolution in galaxies and the enrichment of the intergalactic medium. Their nucleosynthesis yields depend on the supernova type (thermonuclear or core-collapse) and on the mechanism of explosion. However, understanding supernovae remains a challenge for both types. The nature of the progenitor, the explosion mechanism, the ignition of the burning and the propagation of the burning front are still open questions. Large efforts are being made through numerical simulations to pinpoint the mechanism at work, recently emphasizing the role of instabilities and asymmetries in the process. To constrain the models of type Ia (e.g., deflagration, delayed detonation) and core-collapse supernova explosion, the dynamics, the quantity and the three-dimensional distribution of the various synthesized elements in the ejected material are essential observational characteristics to determine the yields and the geometry of the explosion. Young Galactic supernova remnants are key emission sources that can provide these constraints through X-ray observations of the shocked ejecta.

Thanks to its spatially-resolved high-resolution spectroscopy capability, *Athena+* will open a new discovery space by measuring the velocity, temperature, ionization state and composition (including rare elements) of individual parcels of ejected material in the line of sight. For the first time, the three-dimensional geometry and composition of hot ejecta will be revealed in young Galactic supernova remnant, providing unique observational constraints to explosion models of type Ia and core-collapse supernovae.

The *Athena+* also opens new exploring windows on prime SNR issues: shock physics and particle acceleration. The challenge is to understand how particles are accelerated to very high energy at SNR shocks, how they escape and impact their environment. Spatially-resolved high-resolution spectroscopy allows the measurement of ion post-shock temperature, a key information notably for determining the efficiency of particle acceleration. Moreover, access to recombination features of the overionized plasma and charge exchange emission provides a direct path to diagnose the interaction of the supernova with its circumstellar environment and with the cold interstellar medium phase.

Supernovae explosions together with stellar winds are the major players in releasing gas and dust in the interstellar environment. Cold and hot matter coexists in our Galaxy and it is now clear that X-rays are a powerful tool to investigate these different phases of the interstellar medium. The current generation of X-ray observatories revealed the spectroscopic features in emission and absorption of the hot phase in our Galaxy. The imaging of external galaxies has revealed the complex morphology of the hot gas structures showing asymmetries, plumes and clumps, which may be linked to the evolution of the galaxy (e.g. post star-burst episodes). High-resolution spectroscopy offered by the grating instruments revealed also the complexity of the cold phase, via the study of fine structures near the photoelectric absorption edges, caused by dust, and the edges and resonant lines, produced by a multi-temperature gas. This opened up a new opportunity of determining temperatures, abundances, dust depletion and chemistry through X-ray spectroscopy.

Thanks to the synergy between its spatial and spectral resolution, coupled with a large effective area, *Athena+* will provide access to a totally unexplored parameter space. In particular: 1) Reveal temperatures and ionization mechanisms of the complex hot interstellar medium components in our Galaxy as well as in nearby galaxies, through spatially-resolved high-resolution spectroscopy of both emission and absorption features. 2) Scrutinize the dense interstellar medium regions, including the Galactic Centre, for the constituents of dust grains through absorption by Mg, Si and Fe. In these regions the extinction is too high for observations through other spectral windows; 3) Obtain direct measurements of the Fe abundance and inclusion in dust in rather opaque regions ($N_H > 8 \times 10^{22}$ cm$^{-2}$), which are only accessible through the Fe K edge, at 7.1 keV.

## 2. INTRODUCTION

Supernovae (SNe) are the dominant source of heavy elements in the Universe, and they are the only source of alpha-elements (O, Ne, Mg, Si, S, Ar, and Ca) and Fe-group elements (mainly Fe and Ni). The release and subsequent condensation of these elements in the surroundings are believed to have, together with giant-stars mass loss, a major role in the interstellar dust creation. SNe explosions are also the most important source of energy input into the interstellar medium (ISM), both in the form of kinetic/thermal energy and in the form of cosmic rays. Therefore they provide a source of heating of the diffuse gas. Hot gas is indeed found distributed in the Galactic plane, in large-scale structures in the surrounding of the Galaxy, and in the form of plumes above the disk, the latter suggesting a direct link with recent SNe explosions.

## 3. EXPLOSION MECHANISM OF SUPERNOVAE AND PRODUCTION OF HEAVY ELEMENTS

Dynamics, ionization state and composition of galactic supernova remnants (SNR) are the essential characteristics that can help to constrain the explosion mechanism of type Ia and core-collapse supernovae, to quantify their associated production of heavy elements and their impact on the ISM. For the first time, *Athena+* will allow the analysis of individual parcels of ejected material along the line of sight and determination of the velocity, temperature, ionization state and composition.

Understanding the mechanism of explosion of thermonuclear or type Ia supernovae (SN Ia) and core-collapse supernovae (SNcc) remains by itself a challenge, which has a direct impact on our understanding of the chemical evolution in galaxies and the enrichment of the intergalactic medium. Large efforts have been made though numerical simulations to pinpoint the mechanism at work and the properties of the progenitors. A key issue is the necessity of a three-dimensional configuration and treatment of the physical processes at play. X-rays play an indispensable role in these issues as they provide full access to the shocked ejecta, which corresponds to most of the observable mass in supernova remnants. The remaining un-shocked material is, in its largest fraction, invisible at any wavelength. Information on the fraction of the ejecta that does emit at low temperature, can be obtained through optical and IR observations.

SN Ia are the main providers of iron (~75%) and are in addition used as standard candles for cosmology. Understanding their explosion mechanism and evolutionary path is a fundamental key issue with far reaching consequences in both astrophysics and contemporary cosmology. They are thought to be exploding C/O white dwarfs in a binary system. However, a major uncertainty remains on the nature of the companion and the evolutionary paths, which lead to their explosion: either accretion of material from a normal star (single degenerate scenario), or merging with another white dwarf (double degenerate scenario). Regarding the mechanism, the main issue concerns the ignition of the burning and the propagation of the burning front: detonation versus pure deflagration. The production of heavy elements depends directly on these issues through the accretion rate from the companion and the physics of the propagating flame fronts. The presence of an excess amount of Fe or $^{44}$Ti, in the outer ejecta layers can be used to investigate the viability of sub-Chandrasekhar models. One of the recent advances in this field was the detection of the rare elements Cr and Mn with Suzaku (Tamagawa et al. 2009). Uneven mass number elements like Mn in SN Ia are directly linked to the metallicity of the progenitor star and can be used to establish whether a SN Ia originates from a Population I or II star (Badenes et al. 2008b). Recent theoretical studies emphasize the importance of SN ejecta asymmetries, which can be a result of the explosion process itself (Maeda et al. 2010), with a significant impact on the yields, or of the shielding of ejecta by a nearby donor star (Marietta et al. 2000). Statistical analysis of X-ray data showed that the remnants of SN Ia are geometrically distinct from those of SNcc (Lopez et al. 2009). Modeling the global X-ray spectra of SN Ia remnants, like Tycho and 0509-69 (Badenes et al., 2006, Badenes et al. 2008a), has already enabled the exclusion of several SN Ia explosion models. The addition of Doppler velocity and other high-resolution spectral information from *Athena+* opens a new way of investigating the kinematics of SNRs (Figure 1), and it provides 3D determination of the ionization state and abundance. Such observations will place unique constraints on the theoretical models for the explosion mechanism of SN Ia: deflagration versus delayed detonation, level of asymmetry of the explosion through the kinematics of various abundant elements (Si, S, Fe) and production of rare elements (e.g., Mn, Cr).

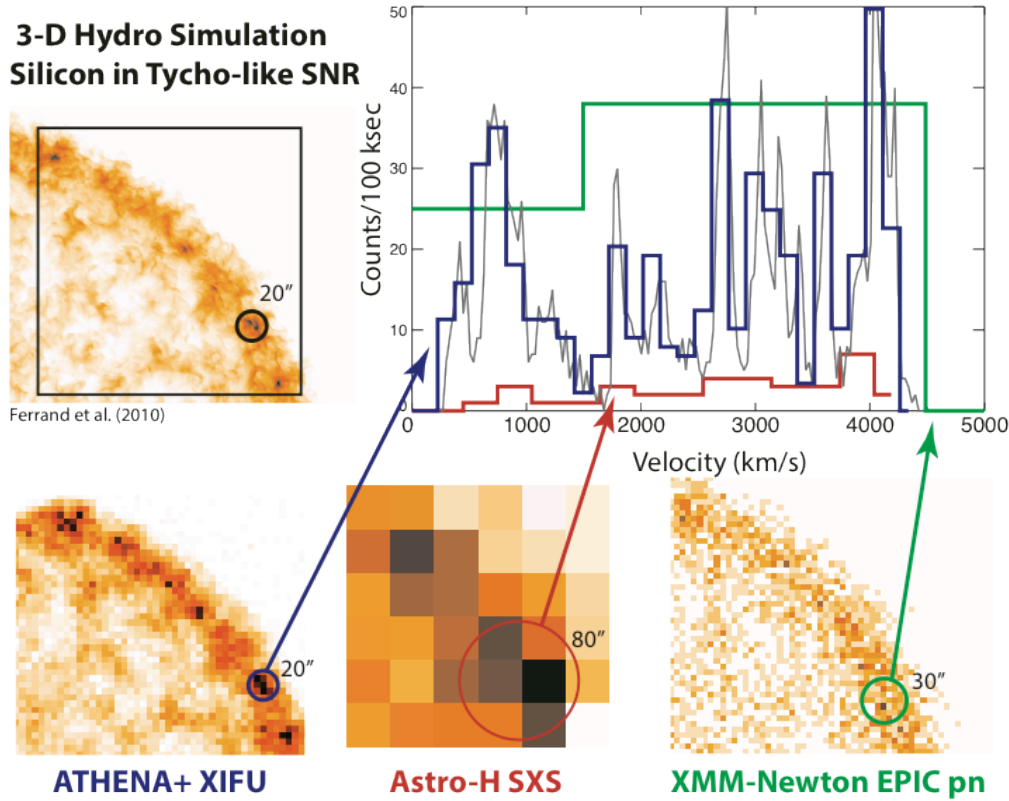

**Figure 1.** *Athena+* X-IFU, *ASTRO-H*, and *XMM-Newton* images and silicon velocity profiles based on a 3D hydrodynamic simulation of Tycho's supernova remnant (Ferrand et al. 2012, 2010) are shown with the typical imaging resolution of the instruments (circles). The *Athena+* X-IFU profile is shown in blue, *Astro-H* SXS in red and *XMM-Newton* in green. Only *Athena+* X-IFU has sufficient spectral and angular resolution, and sensitivity to isolate the highlighted knot and retrieve the velocity information that will reveal the 3D dynamics of the supernova remnant, providing constraints on the explosion mechanism through the measurements of asymmetries. Together with accurate temperature measurements, the shock velocity will be used to quantify the supernova remnant's ability to accelerate cosmic rays.

SNcc are the main providers of intermediate elements (Si, S, and Ca) and are responsible for the enrichment of the early Universe. They explode because of the collapse of the core of a massive star into a neutron star or black hole. However, the value of the mass-cut between the residual compact object and the ejected envelope is poorly known in current theories. The release of gravitational energy powers the explosion of the rest of the star, but how the collapse energy results in the explosion is also an issue (neutrino-driven, MHD-driven) and the process has been difficult to recreate with computer models. Neutrino-plasma coupling is thought to be an important ingredient for the explosion and the instabilities produced during the infall of plasma onto the core should facilitate the explosion (e.g. Blondin et al. 2003). The association of gamma-ray bursts with core collapse SNe has revived the discussion on whether bipolarity or even jet formation are important for all core collapse SNe (Wheeler et al. 2002). Neutrinos can convert protons into neutrons, providing a nucleosynthesis path to rare elements, whose (non)detection holds important clues on the role of neutrinos in SNcc explosions (Wanajo et al. 2011). The instrument for these studies will be *Athena+* X-IFU that will enable us to map out the 3-D structure of Galactic SNRs and provide the 3-D geometrical properties of SN explosions to link them to specific explosion models. The third dimension will be obtained through accurate Doppler shift measurements, allowing the determination of the spatial distribution of the ejected knots, their kinematics, temperature, ionization state and composition accurately mapping ejecta abundance patterns, both in the plane of the sky and in the line of sight. Having access to different groups of synthesized elements provide complementary constraints on SNcc. The quantity of elements lighter than Si (essentially produced during hydrostatic evolution and spread away in the explosion) constrains the nature of the progenitor. The intermediate elements (enhanced through explosive oxygen burning) are related to the explosion energy and the amount of accreted matter onto the core before the explosion. The Fe-group nuclei are indispensable to set constraints on the explosion and mass-cut. Finally, the detection of rare elements in Fe-rich ejecta will constrain various processes affecting the nucleosynthesis processes in the core, in particular those involving neutrinos.

## 4. SHOCK HEATING AND PARTICLE ACCELERATION

Shock physics in supernova remnants has still a number of open issues, notably related to the sharing of the energy at the shock between the different components: thermal plasma (electrons and ions), accelerated particles and magnetic fields.

The magnetic field is a principal player in the diffusive shock acceleration, but also in the level of heating of electrons and therefore the level of temperature equilibration between ions and electrons. Using thermal Doppler broadening, *Athena+* X-IFU will open the possibility to perform direct measurements of post-shock ion temperature along the shock in X-rays. In order to separate thermal line broadening from bulk motions, *Athena+* X-IFU will be used, as it provides the spatial resolution to isolate regions close to the SNR edge, as well as the spectral resolution sufficient to measure line broadening of a few hundreds of km/s. Together with the measured electron temperature, this will determine the level of temperature equilibration between ions and electrons along the shock. These are key measurements, notably to explore cosmic-ray acceleration properties. It has been also noted that efficient cosmic-ray acceleration gives rise to plasma temperatures that are much lower than expected from the standard Hugoniot shock-relations (Decourchelle et al. 2000, Vink et al. 2010). Although low spectral resolution X-ray spectroscopy can be used to measure electron temperatures, the energy in the thermal plasma is characterized by the ion temperatures, which may differ dramatically from the electron temperatures.

Cosmic rays with energies of $10^9$-$10^{15}$ eV are thought to originate mainly from SNR shocks. While radio synchrotron emission of SNR has proven the acceleration of electrons up to GeV energies, X-ray synchrotron emission provided the first evidence for the presence of electrons with energies of $10^{13}$-$10^{14}$ eV at SNR shocks (Koyama et al. 1995). X-ray astronomy is indispensable to understand cosmic-ray acceleration through the detection and characterization of X-ray synchrotron emission from SNRs, which arises from the highest-energy accelerated electrons. Since these high-energy electrons lose their energy on time scales of several years, their location near the shock and the narrowness of the synchrotron regions indicate high amplified magnetic fields (e.g. Vink & Laming 2003), probably itself a result of efficient cosmic-ray acceleration (Bell 2004). The exact shape of the synchrotron spectral cut-off (which falls in the X-ray domain for young supernova remnants, Reynolds & Keohane 1999) depends on the physical mechanisms that limit the energy achieved by the electrons in the acceleration process (e.g. radiative losses, limited acceleration time available, change in the availability of MHD waves above some wavelength) and by the nature of the diffusion process (e.g. Bohm diffusion, Kolmogorov diffusion, etc). The high effective area provided by the *Athena+* WFI camera will allow us to study the high-energy tail of the electrons accelerated at the shock front to discriminate between these different scenarios. In addition, *Athena+* X-IFU will make possible the detection of weak thermal emission in non-thermal rims, whose characteristics holds information, amongst others, on the acceleration efficiency through back-reaction of cosmic rays on the properties of the thermal gas. Finally, *Athena+* WFI will be indispensable to provide identification and characterization of the GeV-TeV gamma-ray sources that will be revealed by the CTA (Cherenkov Telescope Array).

## 5. SUPERNOVA REMNANTS AND THE HOT INTERSTELLAR MEDIUM

The hot phase of the ISM is present in the disk and in the diffuse halo of both our Galaxy and nearby galaxies (Wang et al. 2010 and references therein) in the form of diffuse gas, plumes, as well as high-velocity cloud structures (e.g. Sembach et al. 2003). It is an important tracer of star formation as the gas is kept hot (T=$10^{6-7}$ K) mainly by stellar feedback, i.e. both by young stars and supernovae. In particular the interaction of core-collapse SNRs with their complex environment (wind-blown bubbles and cavity walls and, in general, the complex molecular cloud where the progenitor star was born) drives the transfer of mass and energy from the SNRs to the ISM.

The charge exchange (CX) X-ray emission is a powerful tracer for both the locations where SNR hot gas interacts with neutral gas (notably in the thin post-shock layer) and the interfaces between the hot and cold gas in the ISM (Lallement 2004). This is key information to understand the feedback processes of supernovae, starburst galaxies and galactic winds. CX emission is made exclusively of lines, and high-spectral resolution is mandatory to assess their presence and importance. *Athena+* X-IFU provides the vital capabilities to explore this field. There have been a number of observed X-ray spectral anomalies or unexpected emissions attributed to the CX process during the recent years in supernova remnants (e.g., Cygnus Loop: Katsuda et al. 2011, Puppis A: Katsuda et al. 2012), in a star forming region (Carina: Townsley et al. 2011) and in starburst galaxies and galactic winds (e.g., Kaneda et al. 2010, Liu et al.

2012). High-velocity clouds in the halo may also be the source of CX X-rays (Lallement 2004). All these objects will be studied with *Athena+* X-IFU. Their spectra can be directly diagnosed for CX process, through for instance stronger-than-expected forbidden lines in helium-like triplets, high n-shell enhancements, anomalous abundances or line-to-continuum ratios, and tight correlations with H-alpha emission.

A central enhancement with the presence of overionized plasma has recently been revealed in several SNRs. This corresponds to a recombining plasma, which underwent a rapid cooling, and has a higher ionization stage than that at equilibrium. This is believed to be the result of an early expansion in the dense wind of the red-giant-branch phase of the progenitor. This phenomenon is related to cosmic-ray acceleration, because a very young SNR (fast shock at 10,000 km/s) in a dense wind is one of the best place to reach the knee energy ($3 \times 10^{15}$ eV) in the cosmic-ray spectrum. This overionized plasma is characterized by radiative recombination features in the spectrum. It will be straightforward to measure the prevalence of such features with the excellent spectral resolution of the *Athena+* X-IFU, therefore testing the above scenario.

The connection between the hot and cold phase, crucial for Galaxy formation and evolution theories, has never been investigated in X-rays so far, as the spectral effect of CX (and recombination) features are in general extremely difficult to observe with the current generation of instruments. This science window will be only accessed through *Athena+* X-IFU.

The older population of supernova remnants provides the path to investigate supernovae impact on their environment and the overall building of the hot phase of the ISM. Older supernova remnants possess cooler temperatures ($< 10^7$ K) and their spectra are dominated by a forest of lines around and below 1 keV. That forest could not be resolved by CCDs and the estimates of the gas parameters (temperature, ionization state, abundances, and emission measure) had to rely on coronal plasma models and global spectral fitting. Only a few bright SNRs in the Magellanic Clouds have the right size to be observed at high spectral resolution (e.g. Behar et al. 2001) with the gratings on *XMM-Newton* or *Chandra*. The spatially-resolved spectroscopic capability of *Athena+* X-IFU will enable old SNRs from our Galaxy to be observed with high-spectral resolution better, down to 0.5 keV, than that afforded for global spectra of a few remnants in the Magellanic Clouds by current gratings. Old SNRs are also important agents of dust destruction, via grain sputtering behind the shock. This should show up as a progressive release of refractory elements from the grains to the gas, leading to increasing line emission, different from one element to another. The length scale over which this occurs is a good measure of the efficiency of the process. At present it is very difficult to test because the lines of various elements are not well separated, but the spatial and spectral resolution of *Athena+* X-IFU are adequate to measure the effect in local SNRs.

Population studies cannot be conducted in our Galaxy, due to uncertain distances and very different absorption along different lines of sight. The effective area of *Athena+* X-IFU is about 100 times larger than that of the *XMM-Newton* RGS so it can study the population of LMC and SMC SNRs in great detail, but also access SNRs in M31 and M33. *Athena+* will allow us to study these explosions, their products and their immediate surroundings, thereby providing an important baseline for studies of metal abundances throughout cosmic time (see Ettori, Pratt et al. *Athena+* supporting paper).

SN activity is a major source of ionization for the hot phase of the ISM. With the current generation high-resolution instruments, only the sight lines towards bright targets could be thoroughly investigated (Juett et al. 2004, Yao & Wang 2006). With the *Athena+* large effective area and energy resolution the mapping of the general ISM hot gas phase, as seen in absorption against an impressive number of X-ray sources will be possible. Indeed, the high effective area will enable access to more than 1800 sources with flux larger than $10^{-12}$ erg/cm$^2$/s located in and around the Galactic disk. This will allow us to characterize the temperature distribution of the multiphase hot ISM, revealing the different ionization processes in action (e.g. collisional excitation by shocks, photoionization by X-ray sources). The combined study of emission of an empty field, and absorption against an X-ray source of the hot ISM along adjacent lines of sight (e.g. Hagihara et al. 2011), will reveal the structures and the density of the ionized gas. This will be only possible thanks to the capability of the *Athena+* X-IFU.

In parallel, spatially-resolved spectroscopy of emission both in our and external galaxies can reveal the morphology and local characteristics of the hot ISM. Indeed, this gas whose distribution is often seen to be asymmetric and clumpy (e.g. Strickland et al. 2004, Li & Wang 2007), is likely to keep the echo of past starburst episodes (Tang et al. 2009) of a recent merging or an active galactic nuclei activity (see Cappi, Done et al. *Athena+* supporting paper), often characterized by giant galactic outflows (Diehl & Statler 2008). Spatially-resolved spectroscopy, which will only be possible with *Athena+* X-IFU, of those plumes and clumps will again reveal the nature of the ionizing source.

# 6. THE COLD ISM PHASE

The ISM is a dynamical environment (e.g. Ferriere et al. 2001) where a cold dust and gas phase (T < 5000 K) coexist with the warm/hot gas phase (Spitzer 1990). The diffuse cold gas and dust follow in general the shape of the spiral arms, displaying a higher column density in the central few pc of the arm (e.g. Dame et al. 2001, Kalberla & Kerp 2009). In the Galactic Center region, the column density is on average larger also due to a larger number of molecular clouds along our line of sight (Dame et al. 2001). The distribution of the cold matter is not homogenous, but it rather shows filaments and local substructures which suggest a constant interplay between the stars and the surrounding (e.g. Kalberla & Kerp 2009). The cold matter interacts with X-ray photons from background sources (e.g. X-ray binaries) absorbing and scattering the radiation. The X-ray band offers unique advantages with respect to longer wavelength studies in this field. First, an energy window where absorption by dust grains containing O, Mg, Si and Fe, which are the main constituents of silicates, is represented (e.g. Lee et al. 2009, Costantini et al. 2012, Pinto et al. 2013). This implies the access to two Fe features in the X-ray band. Fe features are elusive at other wavelengths, while determining Fe inclusion in dust is fundamental in any dust models. X-rays are moreover sensitive to a large range of column densities ($\sim 10^{20}$-$10^{23}$ cm$^{-2}$). Finally, X-ray measurements allow a simultaneous study of the two constituents of extinction: scattering and absorption, which are sensitive to different dust properties.

Through the absorption features imprinted by dust, the chemistry of the grains can be directly accessed. The interaction of the photoelectric wave inside the dust grain produces fine-structure features, which can be few electronvolt wide (Figure 2). From the depth and the number of such structures, the nature of the absorbing agent can be derived. Current instrumentations could only focus on the soft X-ray features (e.g. O K, Fe L edge) which are representative of a tenuous medium. *Athena+* X-IFU instead will access with unprecedented energy resolution the region beyond 1 keV ($\lambda$<12.4 Å), where the edges of Mg, Si, and Fe are present (Figure 2). These edges of highly depleted elements are representative of dense interstellar matter environments (e.g. near the Galactic Centre). These regions are inaccessible at longer wavelengths, due to the heavy extinction. The *Athena+* X-IFU will be the only instrument able to study in detail the Fe K edge at 7.1 keV, which will be enhanced only for extremely high column densities ($N_H > 8 \times 10^{22}$ cm$^{-2}$, Figure 2). With *Athena+*, high-quality spectra not only of the brightest sources, but also of a fainter sources population will be collected. This ensures an accurate mapping of the Milky Way, measuring the chemical composition of different sight lines. Again thanks to the high throughput of *Athena+*, this study will be extended for the first time to nearby galaxies, as spectroscopy of the cold ISM, seen against bright X-ray sources will be possible. This will open up a new window on the dust properties in extragalactic environments as seen in the X-rays.

Scattering of X-rays by dust particles is studied through direct imaging of the halo of diffuse emission caused by dust grains. Halos are mainly sensitive to the grain size distribution and geometrical distribution along the line of sight (e.g. Predehl & Schmitt 1995; Smith et al. 2006, Valencic et al. 2009). Since the scattered radiation is a small fraction of the light of the background source, the study of the halo profiles has been restricted to the brightest sources. With *Athena+* WFI, faint and extended scattering halos profiles, which may for example be produced by the smaller grain particles population (e.g. Draine 2003), will be for the first time routinely studied. This will allow combining at the same time information from the X-ray, optical and infrared domains on dust along the same line of sight. Thanks to the capabilities of *Athena+* X-IFU, spatially-resolved spectroscopy of the scattering halo, at a given scattering angle, can be performed. This measurement is sensitive to the chemistry of dust only (Predehl & Klose 2006) and it has been attempted so far only on an exceptionally bright source (Costantini et al. 2005). With *Athena+* spatially-resolved spectroscopy of halos around a number of bright sources ($F_{(2-10\ keV)} \sim 10^{-8}$-$10^{-9}$ erg cm$^{-2}$ s$^{-1}$) will be successfully performed with moderate exposure times. This will allow a parallel analysis of the dust chemistry using both scattering and absorption. Scattering halos are also a powerful tool for distance determination provided that the background source is variable (e.g. eclipsing binaries, Predehl et al. 2000; gamma-ray bursts, Vaughan et al. 2004; magnetars, Tiengo et al. 2010). This method is purely based on the fact that after a flux variation of the source, soft X-ray photons, which are efficiently scattered, will be observed with a delay with respect to the hard, unscattered, ones. Such delay is a function of the distance of the source and the dust distribution along the line of sight. Individual dust clouds in our Galaxy can be studied with greater detail if a bright sudden burst from a background X-ray source is observed through them. Expanding X-ray rings are observed from which both the cloud distance and dust grain size distribution can be precisely determined (e.g. Tiengo et al. 2010). The characteristics of *Athena+* WFI, coupled with a fast re-pointing of the observatory, will allow us to probe a large number of Galactic clouds. The study of dust particles through absorption and scattering in X-rays, using the *Athena+* radically-improved instrumental capabilities, will be fundamental for any global multiwavelength modeling of the chemical and physical properties of the dust grains.

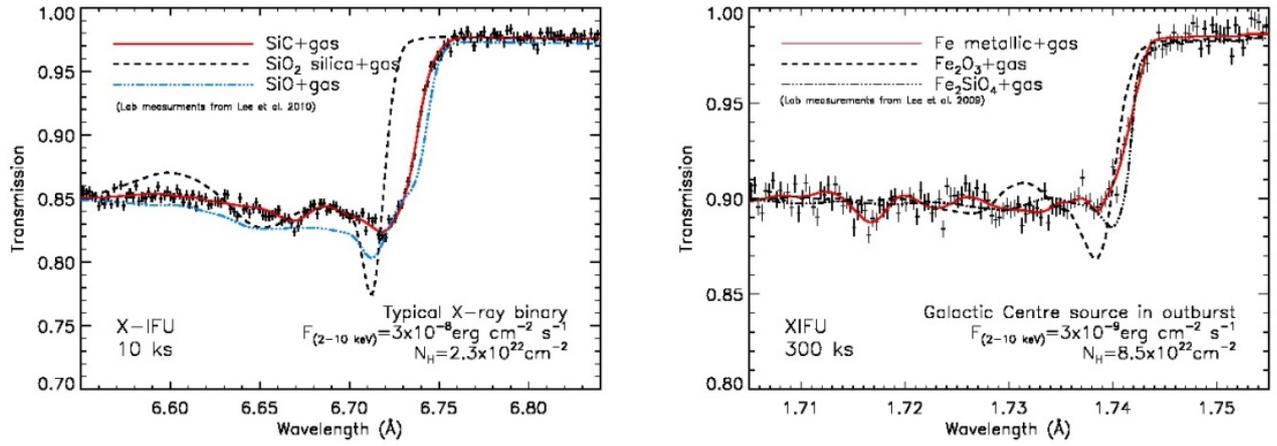

**Figure 2:** The *Athena+* X-IFU will access with unprecedented instrumental capabilities the region beyond 1 keV (λ<12.4 Å). The edges of Mg and Si (left panel) will be easily studied, providing an insight into the chemistry of interstellar dust in denser environments. The *Athena+* X-IFU will be the only instrument capable of studying the iron inclusion in dust in the densest regions of our Galaxy. Such large extinction regions will enhance the Fe K edge (right panel). Both panels show *Athena+* X-IFU data compared against different mixtures of gas and dust.